\documentclass[twocolumn,groupedaddress,
superscriptaddress,
showpacs,preprintnumbers,
amsmath,amssymb,
aps,pra,
longbibliography
]{revtex4-1}

\usepackage{graphicx}
\usepackage{dcolumn}
\usepackage{bm}%
\usepackage[colorlinks=true,citecolor=blue]{hyperref}
\hypersetup{colorlinks=true,citecolor=blue,linkcolor=red,urlcolor=blue}

\usepackage{float}

\usepackage{color}
\definecolor{Green}{RGB}{0,204,102}
\definecolor{Purple}{RGB}{102,0,255}
\definecolor{Blue}{RGB}{51,153,255}
\definecolor{Red}{RGB}{151,010,010}

\begin{document}

\sloppy

\title{Hyperbolic plasmon modes in tilted Dirac cone phases of borophene}

\newcommand*{\DIPC}[0]{{
Donostia International Physics Center (DIPC),
Paseo Manuel de Lardizabal 4, 20018 Donostia-San Sebasti\'an, Spain}}

\newcommand*{\IFS}[0]{{
Institute of Physics, Bijeni\v{c}ka 46, 10000 Zagreb, Croatia}}

\newcommand*{\zahra}[0]{{
School of Nano Science, Institute for Research in Fundamental Sciences (IPM), Tehran 19395-5531, Iran}}

\newcommand*{\rezaa}[0]{{
School of Physics, Institute for Research in Fundamental Sciences (IPM), Tehran 19395-5531, Iran}}

\newcommand*{\rezab}[0]{{
School  of  Physics,  University  of  New  South  Wales,  Kensington,  NSW  2052,  Australia}}

\newcommand*{\rezac}[0]{{
ARC Centre of Excellence in Future Low-Energy Electronics Technologies, UNSW Node, Sydney 2052, Australia}}

\author{Zahra Torbatian}
\affiliation{\zahra}
\author{Dino Novko}
\email{dino.novko@gmail.com}
\affiliation{\IFS}
\affiliation{\DIPC}
\author{Reza Asgari}
\email{asgari@ipm.ir}
\affiliation{\rezaa}
\affiliation{\rezab}
\affiliation{\rezac}

\begin{abstract}
Hyperbolic materials are receiving significant attention due to their ability to support electromagnetic fields with arbitrarily high momenta and, hence, to achieve very strong light confinement. Here, based on first-principles calculations and many-body perturbation theory, we explore the characteristic of two-dimensional plasmon modes and its hyperbolic properties for two phases of single layer boron hosting tilted Dirac cone, namely, the {\it hr}-{\it sB} and 8{\it Pmmn} borophene. In-plane anisotropy in borophene is manifested in the structural, electronic, vibrational and optical properties. We find two hyperbolic regimes for both phases of borophene, where the high-energy one is located in the visible range. The {\it hr}-{\it sB} borophene is characterised with an intrinsic high carrier density and it supports strong hyperbolic plasmon modes in the visible part of the spectrum. The 8{\it Pmmn} borophene, on the other hand, resembles the prototypical Dirac material graphene, and upon carrier doping acquires anisotropic Dirac plasmons in the mid-infrared. We have also investigated the impact of the electron-phonon coupling and Landau damping on these hyperbolic plasmon modes. Our results show that borophene, having high anisotropy, intrinsic high carrier concentration, low-loss hyperbolic Dirac plasmon modes, and high confinement can represent a promising candidate for low-loss broad band surface plasmon polariton devices.
\end{abstract}


\maketitle

\section*{Introduction}\label{sec:intro}
Natural hyperbolic materials with anisotropic dielectric constants where the real parts of the principle components have opposite signs, have recently attracted many attention\,\cite{poddubny2013hyperbolic,nathyper14,gjerding2017layered,biswas2021tunable,he2021guided,chen2020flat}. These materials support electromagnetic modes with large wave-vector and exhibit directionally propagating polaritons with extremely large photonic density of states\,\cite{caldwell2014sub,nathyper14,hyperBP3,Ma2018}. Polaritons in natural materials can be formed when the light interacts with the intrinsic plasmons, phonons, or excitons\,\cite{low16,basov20}.
In contrast to the artificial metasurfaces, where hyperbolic dispersion is limited to wave-vectors smaller than the inverse of the structure size, natural hyperbolic materials support high electromagnetic confinement and contain no internal interfaces for the electrons to scatter off\,\cite{nathyper14}. Naturally occurring hyperbolic plasmonic media are rare and thus far they were experimentally verified to exist only in the WTe$_2$ thin film\,\cite{wang2020van,wte2exp2}. However, recent theoretical studies have predicted interesting hyperbolic properties not only in WTe$_2$\,\cite{torbatian2020tunable}, but as well in black phosphorous\,\cite{hyperBP2,hyperBP3,hyperBP1}, cuprates, MgB$_2$\,\cite{nathyper14}, electrides\,\cite{electrides17}, MoTe$_2$\,\cite{wang2020hyperbolicity,edalati2020mote2}, layered hexagonal crystal structures~\cite{PhysRevB.103.035425} and MoOCl$_2$ \cite{zhao2020highly}. Generally, ideal natural hyperbolic plasmonic material is needed to possess high structural and electronic anisotropy, high carrier density and low plasmonic losses. Here, we propose two phases of borophene with such electronic and optical properties that might be naturally suitable to serve as an efficient hyperbolic plasmonic material.

Recently two-dimensional (2D) boron with high free charge carrier concentration emerged as a perfect 2D metal with extraordinary electric, optical and transport properties\,\cite{huang2017two,xie2020two,yan2020theoretical,hou2020borophene,vishkayi2017current,gao2017prediction,Lyudmyla18,lian2020integrated,haldar2020microscopic,Dereshgi20}.
In fact, borophene, as an intrinsic 2D metal with both high carrier density and high confinement, can be a promising candidate to develop optical devices based on low-loss broad band surface plasmon polaritons\,\cite{lian2020integrated,dereshgi2020anisotropic}. Until now, a number of monolayer boron sheets have been theoretically predicted and three types of borophene fabricated in the experiment, i.e., strip, $\beta_{12}$, and $\chi_3$ borophene\,\cite{mannix2015synthesis,feng2016experimental}.

In this paper, we investigate two phases of borophene, namely, 8{\it Pmmn} and {\it hr}-{\it sB}.
The first one represent a polymorph of borophene which hosts an anisotropic tilted Dirac cone\,\cite{lopez2016electronic,verma2017effect,sadhukhan2017anisotropic}. The Dirac cone was thus far confirmed to exist in the $\beta_{12}$ phase of borophene on Ag(111)\,\cite{feng2017dirac} and indicated in reconstructed borophene polymorph on Ir(111)\,\cite{petrovic21}. The second one, i.e., {\it hr}-{\it sB}, is a newly predicted boron monolayer with two types of Dirac fermions coexisting in the sheet; one type is related to Dirac nodal lines and the other is related to tilted semi-Dirac cones with strong anisotropy.
This new allotrope possesses a high stability compared to the experimentally available $\delta_6$, $\beta_{12}$ and $\chi_3$ sheets\,\cite{zhang2017dirac}.
Here, we investigate characteristics of 2D Dirac plasmons and its hyperbolic properties for these two phases of borophene in the framework of density functional theory and many-body perturbation theories. Our results show that both phases of borophene support strong Dirac hyperbolic plasmons due to their highly anistropic optical properties. Owing to the high intrinsic carrier concentration, hyperbolic Dirac plasmons in the {\it hr}-{\it sB} borophene are present also in the visible range. In addition, we explore the coupling of Dirac plasmons with phonons as well as with the electron-hole pairs from first principles. This allows us to study the decay rates of plasmons in borophene and asses its efficiency both qualitatively and quantitatively. Interestingly, we find extremely negligible plasmonic losses in the 8{\it Pmmn} borophene, which are actually comparable to the losses in prototypical Dirac material graphene. Therefore, these novel 2D hyperbolic Dirac materials are promising and appealing for applications in plasmonics.


\begin{figure}[!t]
\begin{center}
\includegraphics[width=8.0cm]{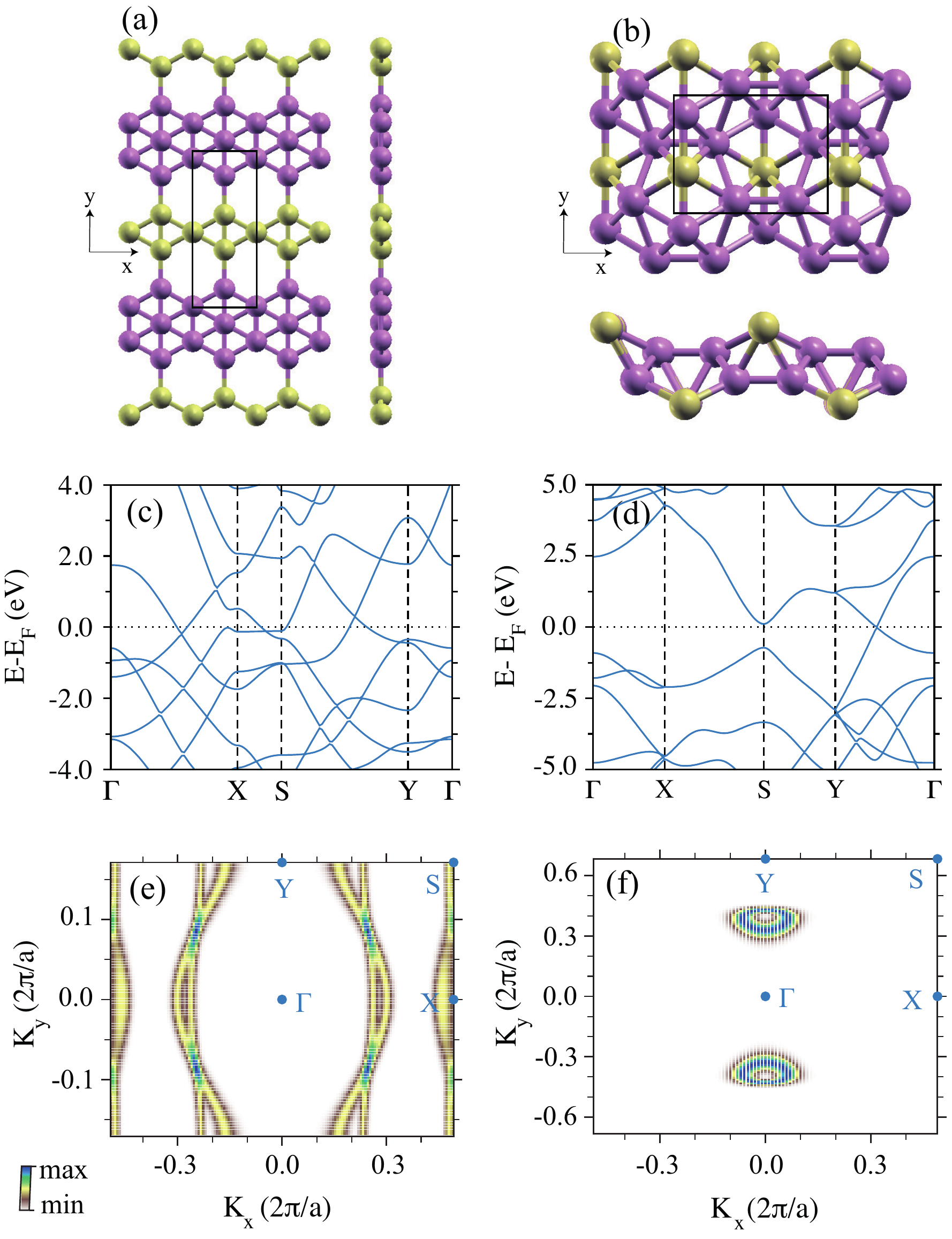}
\caption{{\bf Geometric and electronic structure of borophene.} Top and side views of optimized (a) {\it  hr}-{\it sB} and (b) 8{\it Pmmn} borophene structures. The electronic band structure along high symmetry direction of {\it  hr}-{\it sB} and 8{\it Pmmn} are shown in (c) and (d), respectively. Panels (e) and (f) show the Fermi surfaces of pristine {\it  hr}-{\it sB} and hole doped 8{\it Pmmn} ($E_F=-250$\,meV).
The boron atoms in the hexagon stripes (B$_h$) (inner atoms B$_I$) and in rhombus stripse (B$_r$) (ridge atoms B$_R$) are indicated in purple and yellow, respectively, in the {\it  hr}-{\it sB} (8{\it Pmmn}) phase. The primitive unit cell in both structures is marked by the black rectangle.}
\label{fig1}
\end{center}
\end{figure}

\section*{RESULTS AND DISCUSSION}\label{sec:result}

{\bf Electronic structure of borophene.--} In Fig.\,\ref{fig1} we display the optimized structure and electronic bands along high-symmetry points for the {\it hr}-{\it sB} and 8-{\it Pmmn} phases of borophene. The unit cells of both phases are rectangular and contain eight atoms. In contrast to the 8{\it Pmmn} phase with buckled structure, the {\it hr}-{\it sB} phase is purely planar with stripes of hexagons and rhombs along the $x$ axis [see Fig.\,\ref{fig1}(a)]. The boron atoms are called B$_h$ and B$_r$ in the hexagon and rhomb stripes, respectively.
The 8{\it Pmmn} borophene structure contains two types of non-equivalent B atoms, namely the ridge atoms (B$_R$) and the inner atoms (B$_I$) [see Fig.\,\ref{fig1}(b)], leading to various physical and chemical properties. Figure \ref{fig1}(c) shows the anisotropic band structure of {\it hr}-{\it sB} with a Dirac nodal line and tilted semi-Dirac cones coexisting around the Fermi energy\,\cite{feng2017dirac}. Two linear bands are crossed between $\Gamma$ and X points as well as between S and Y points forming Dirac point slightly bellow and above the Fermi level. Actually, these apparently two structures are part of the Dirac nodal line, which can be seen from the Fermi surface plot in Fig.\,\ref{fig1}(e). The tilted Dirac cone is at the edge of the Brillouin zone (BZ). The orthorhombic 8{\it Pmmn} borophene possesses tilted anisotropic Dirac cone along the $\Gamma$Y path, where the Dirac point crosses the Fermi level [see Fig.\,\ref{fig1}(d) and the corresponding Fermi surface with $E_{\rm F}=-250$\,meV in Fig.\,\ref{fig1}(f)]. It is worth mentioning that the Dirac cone of the {\it 8{\it Pmmn}} borophene emerges from the symmetry of the hexagonal and topologically equivalent lattice to uniaxially strained graphene lattice\,\cite{lopez2016electronic}.

In contrast to graphene where the two crossing linear bands both come from the $p_z$ orbitals, in 2D boron allotropes the $p_z$ orbitals are not the only dominating contribution to the Dirac cone. In the {\it hr}-{\it sB} sheet, the two crossing bands between the $\Gamma$ and X points come from different orbitals originating in various atomic sites, i.e., the $p_z$ orbitals of $B_h$ atoms and the $p_{x/y}$ orbitals of $B_r$ atoms have dominating contributions. Along the XS direction, $p_z$ orbitals of both $B_r$ and $B_h$ atoms are contributing to the Dirac bands. Furthermore, the Dirac cones in 8{\it Pmmn} are arising from the $p_z$ orbitals of the $B_I$ atoms with a much lesser contribution of the $p_x$ orbitals. In this phase the $B_R$ atoms do not contribute to the formation of the electronic states in the vicinity of the Fermi level, which is in line with the previous theoretical work\,\cite{zhang2017dirac}.

\begin{figure}[!t]
\begin{center}
\includegraphics[width=8.4cm]{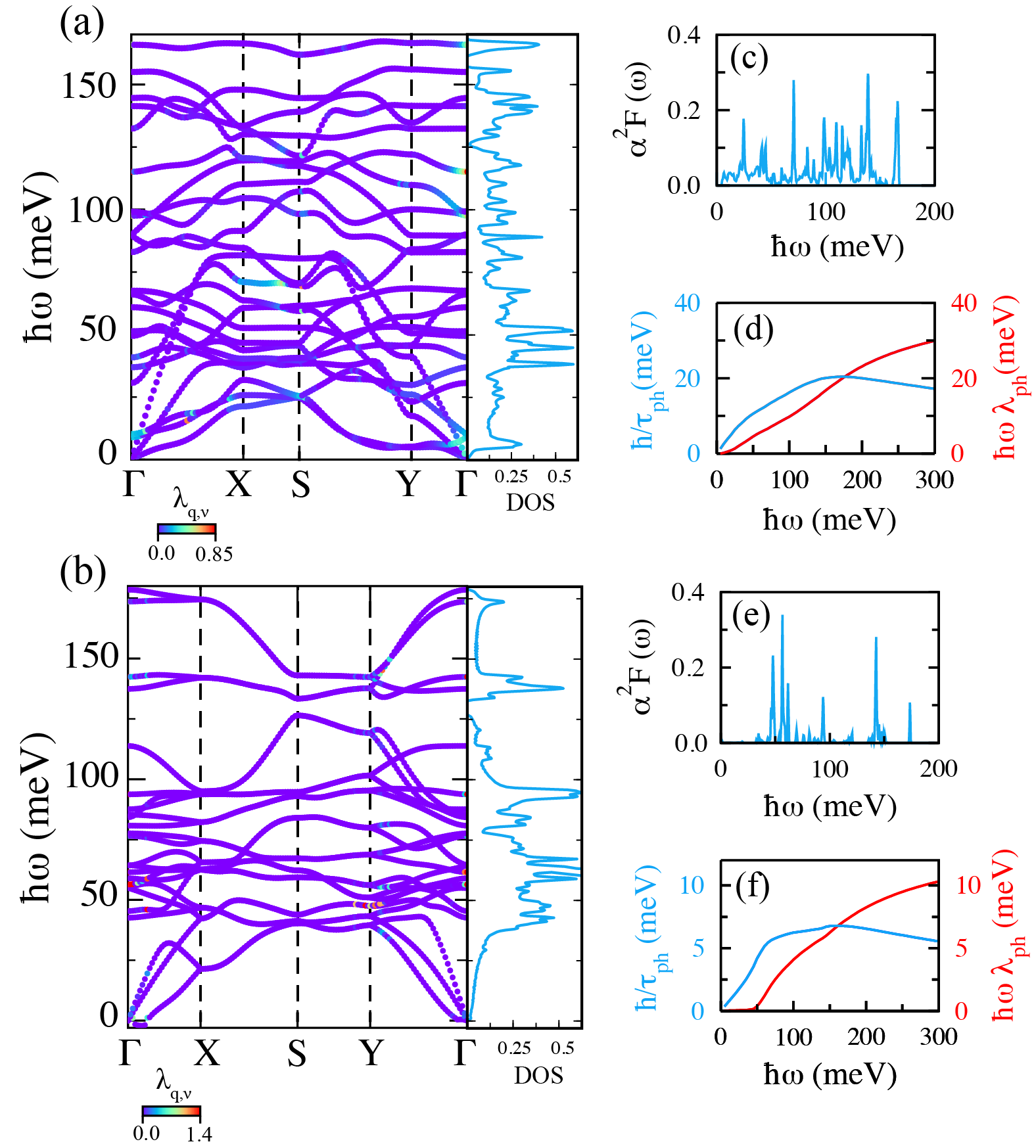}
\caption{{\bf Vibrational properties and electron-phonon coupling in borophene.} The phonon band structure and momentum-resolved $\lambda_{\bf q,\nu}$ along the high symmetry points together with the phonon density of states (DOS) for (a) pristine {\it  hr}-{\it sB} and (b) hole-doped ($E_F=-250$\,meV) 8{\it Pmmn} phases of borophene. The Eliashberg function $\alpha^2F(\omega)$, phonon-induced decay rate $1/\tau_{\rm ph}(\omega)$, and energy renormalization parameter $\omega\lambda_{\rm ph}(\omega)$ are shown in panels (c) and (d) for {\it  hr}-{\it sB}, while for hole-doped 8{\it Pmmn} in panels (e) and (f).
}
\label{fig2}
\end{center}
\end{figure}

The study of the EPC represents a fundamental and fascinating issue and it is the first step towards the calculations of the plasmon decay rates in the two phases of borophene. The dynamical matrices and phonon-perturbed potentials are calculated on a $\Gamma$-centered grid within the framework of density functional perturbation theory. In Figs.\,\ref{fig2}(a) and \ref{fig2}(b), we show the phonon dispersions and momentum-resolved $\lambda_{\bf q,\nu}$ along the high symmetry points (superimposed as a color code to the corresponding dispersions), and phonon density of states (DOS) for two chosen phases of borophene. Note that for the 8{\it Pmmn} borophene, the hole-doped case with $E_{\rm F}=-250$\,meV is considered. No imaginary vibrating mode is seen for the {\it hr}-{\it sB} phase of borophene, indicating the stability of this structure. In 8{\it Pmmn} all the phonon modes are positive frequencies, except the transverse branch near the $\Gamma$ point. Moreover, the results show the EPC strength is remarkable along the XS direction and near the $\Gamma$ point for {\it hr}-{\it sB}. While the EPC strength in 8{\it Pmmn} is weak and appears for the modes at the $\Gamma$ and Y points. The prominent peaks in the phonon DOS of {\it hr}-{\it sB} structure are located around 50, 90, and 150\,meV and are due to the dispersionless nature of the phonon band structure around these energies. In the 8{\it Pmmn} phase the dispersionless phonon structures are also behind the peaks in the phonon DOS.

The Eliashberg function, $\alpha^2F (\omega)$, the
phonon-induced decay rate $1/\tau_{\rm ph} (\omega)$ and energy renormalization parameter $\omega\lambda_{\rm ph}(\omega)$ for the two phases of borophene are illustrated in Figs.\,\ref{fig2}(c)-(f). The feature of the Eliashberg function for the {\it hr}-{\it sB} structure does not follow the phonon DOS for energies below 50\,meV but it resembles the phonon DOS above 50\,meV. The most intense peaks come from the optical phonons near the $\Gamma$ point and along the XS path, where momentum-resolved $\lambda_{\bf q,\nu}$ appears to be the strongest [see Fig.\,\ref{fig2}(a)].
Similarly, the characteristic of the Eliashberg function for 8-{\it Pmmn} are completely unconnected to its phonon DOS and but mainly to its EPC strengths $\lambda_{\bf q,\nu}$ [Fig.\,\ref{fig2}(b)].
Regarding the total EPC strength, we have obtained the values of $\lambda=0.25$ and $\lambda=0.008$ for {\it hr}-{\it sB} and 8{\it Pmmn} phases of borophene, respectively. It is surprising that the EPC in 8{\it Pmmn} is extremely small and it is therefore expected that the corresponding phonon-induced plasmon decay is also quite small.

\begin{figure}[!t]
\begin{center}
\includegraphics[width=8.4cm]{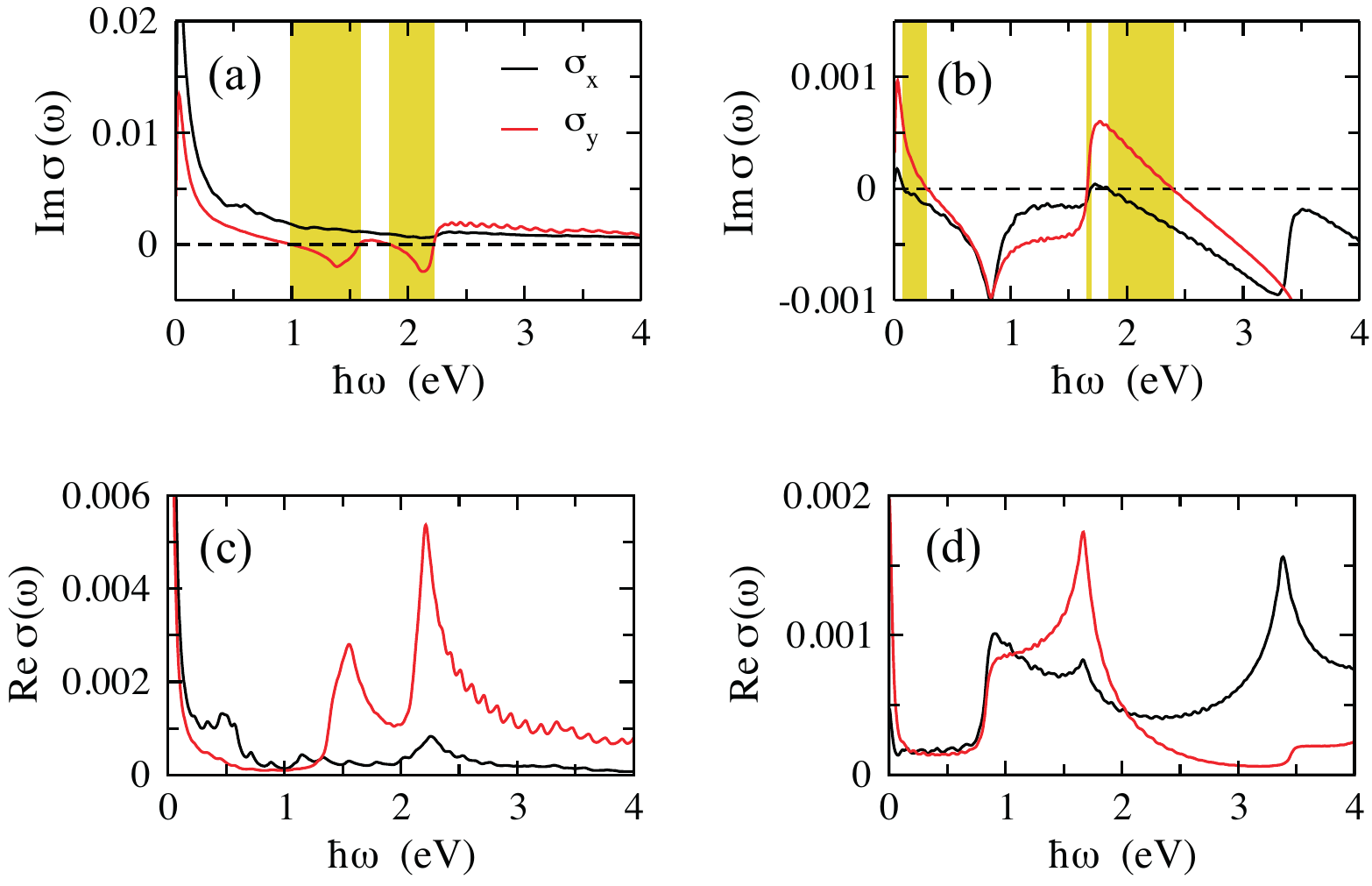}
\caption{{\bf Polarization-dependent optical absorption and hyperbolic regions.} The imaginary (top panel) and real parts (bottom panel) of optical conductivity along the $x$ (black line) and $y$ (red line) directions for (a),(c) {\it  hr}-{\it sB} and (b),(d) 8{\it Pmmn}. The shaded areas show the hyperbolic regions in the both phases of borophene. }
\label{fig3}
\end{center}
\end{figure}

{\bf Optical conductivity and hyperbolic regions.--} We now turn to the study of anisotropic optical properties in borophene.
The appearance of hyperbolic regions for certain photon energy can be inspected by calculating the optical conductivity for different light polarization directions. The corresponding hyperbolic condition for specific photon energy $\hbar\omega$ is defined as
\begin{eqnarray}
\mathrm{Im}\,[\sigma_x(\omega)]\times \mathrm{Im}\,[\sigma_y(\omega)]<0.
\label{hypercond}
\end{eqnarray}
We display the imaginary (top) and the real (bottom) parts of optical conductivity along two principal axes for {\it hr}-{\it sB} (left) and 8{\it Pmmn} (right) phases of borophene in Fig.\,\ref{fig3}. The hyperbolic regimes are depicted with the shaded area. The imaginary part of optical conductivity along the $y$ axis for the {\it hr}-{\it sB} phase changes its sign around 1.0\,eV and stays negative up to 1.58\,eV. This demonstrates the fist hyperbolic region in the {\it hr}-{\it sB} borophene. Interestingly, the second hyperbolic regime falls between 1.84 and 2.21\,eV, i.e., in the visible range. The hyperbolicity condition for the 8{\it Pmmn} phase of borophene is also met in the visible part of the spectrum, i.e., there is one small hyperbolic region around 1.66\,eV, and one with the energy window going from 1.85 to 2.39\,eV. In addition, the 8{\it Pmmn} borophene displays hyperbolic behavior also at the lower energies between 0.085 and 0.27\,eV (mid-infrared).

From the results of the real parts of the optical conductivity, indicating the allowed intraband and interband optical transitions, one can examine the origin of these hyperbolic regions. For instance, in both phases of borophene, there are interband excitations between 1.5 and 2.5\,eV, which are intense and allowed for the $y$ direction, while forbidden and negligible for the $x$ direction. In addition, the 8{\it Pmmn} phase has strongly anisotropic intraband transitions (see the Drude peaks for the $x$ and $y$ directions below 0.2\,eV).
All these results suggest an extremely promising hyperbolic optical property of these two phases of borophene.

\begin{figure}[!t]
\begin{center}
\includegraphics[width=8.4cm]{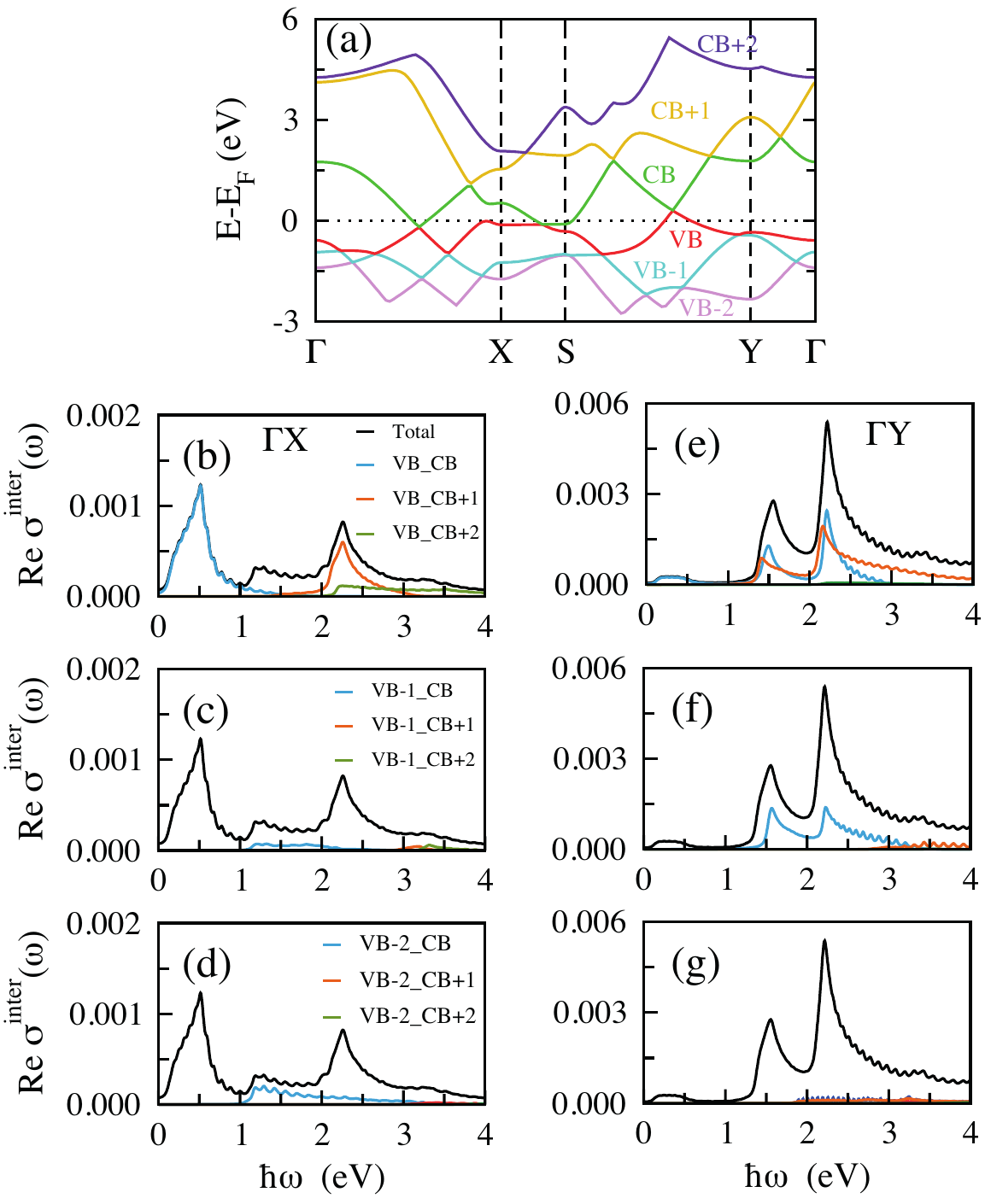}
\caption{{\bf Interband contributions to optical absorption.} (a) The electronic band structure of {\it hr}-{\it sB} where three bands just below (VB$-2$, VB$-1$ and VB) and three bands just above the Fermi level (CB, CB$+1$ and CB$+2$) are illustrated with different colors. The different contributions to the interband optical absorption ${\rm Re}\,\sigma^{\rm inter}_{\mu}(\omega)$ of {\it hr}-{\it sB} coming from the transitions between the bands indicated in panel (a) when the polarization direction is along (b)-(d) $\Gamma$X and along (e)-(g) $\Gamma$Y. Total optical absorption is shown with the black line.}
\label{fig4}
\end{center}
\end{figure}

We further inspect the optical absorption of {\it hr}-{\it sB} that shows intense peaks due to interband transitions at energies below 3\,eV, which can therefore interact with the Dirac plasmons. In Fig.\,\ref{fig4} the real part of the interband optical conductivity ${\rm Re}\,\sigma^{\rm inter}_{\mu}(\omega)$ is shown for particular interband transitions for $x$ and $y$ polarizations, coming from three valence bands (VB$-2$, VB$-1$ and VB) below and three conduction bands (CB, CB$+1$ and CB$+2$) above the Fermi level. The most intense abosrption peaks in the presented energy range stem from the interband transitions from VB to CB, from VB to CB$+1$, as well as from VB$-1$ to CB. We single out three interband excitations relevant for the Dirac plasmon: ($i$) the prominent interband peak at $\sim 0.5$\,eV that is present for the $\Gamma$X direction, ($ii$) absorption peak at $\sim 1.5$\,eV for the $\Gamma$Y direction, and ($iii$) interband continuum that starts at $\sim 1$\,eV for the $\Gamma$X direction. According to Fig.\,\ref{fig4}(b) the interband peak ($i$) comes entierly from the VB to CB transitions around the X point of the BZ. The feature ($ii$) originates from interband transitions around the Y point including VB-1, VB, CB, and CB+1. The continuum ($iii$) is due to interband transitions between VB-2 and CB as well as between VB-1 and CB both around the S point of the BZ.

\begin{figure}[!t]
\begin{center}
\includegraphics[width=7.cm]{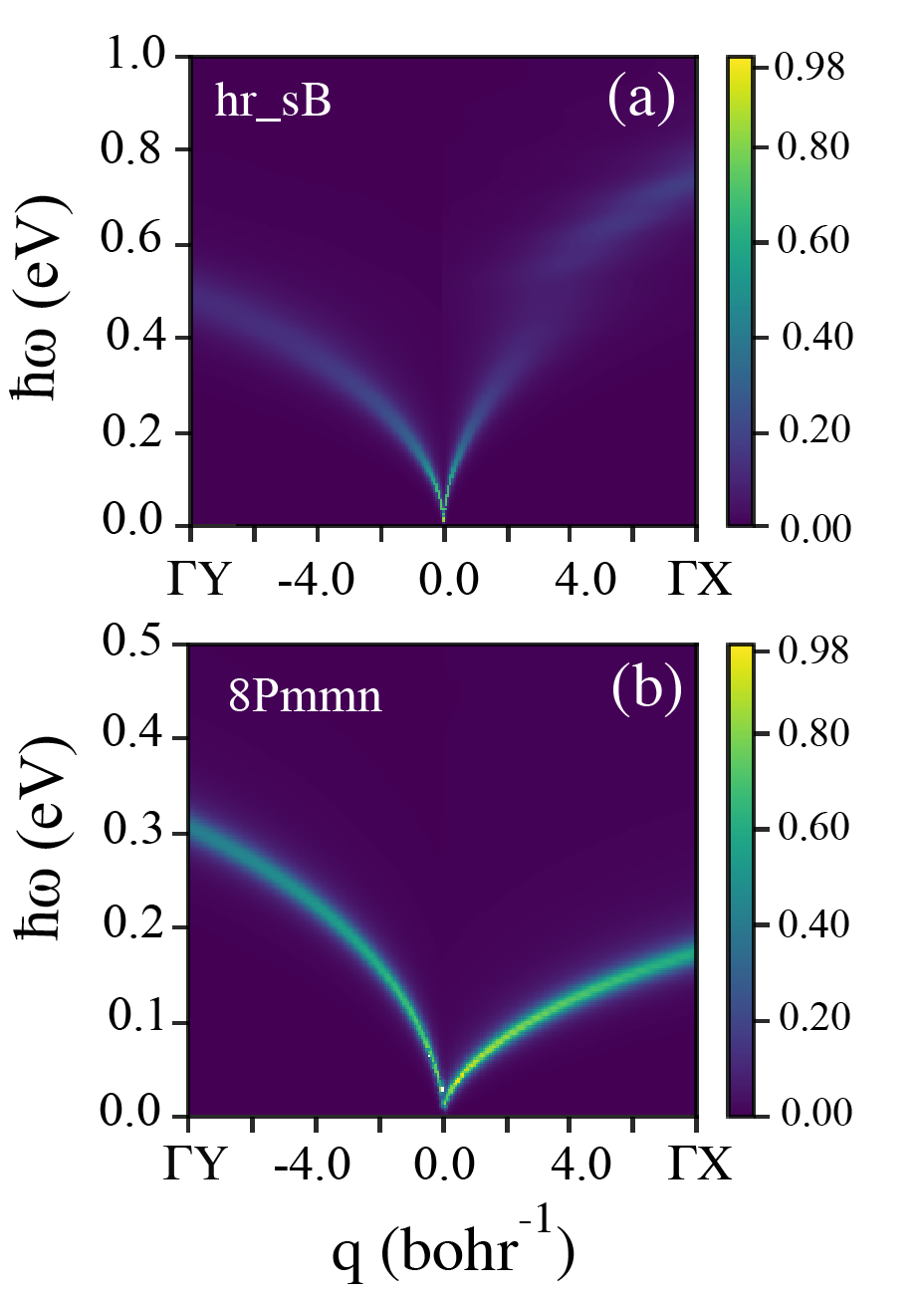}
\caption{{\bf Hyperbolic Dirac plasmons in optical limit.} Plasmon dispersions as a function of $q$ (in 10$^{-3}$ bohr$^{-1}$) with incorporating the electron-phonon coupling at $T=0$\,K for (a) pristine {\it  hr}-{\it sB} and (b) hole doped 8{\it Pmmn} ($E_{\rm F}=-250$\,meV) phase of borophene along two principal axes, i.e., $\Gamma$X and $\Gamma$Y directions.}
\label{fig5}
\end{center}
\end{figure}

{\bf Anisotropic dispersion and damping of Dirac plasmon.--}
In this section, we first study the plasmon dispersion of the {\it hr}-{\it sB} and 8{\it Pmmn} phases of borophene in the optical limit (small ${\bf q}$) for light polarization along the $\Gamma$X and $\Gamma$Y directions, and with including the electron-phonon interaction. To do so, we adopt utilize the current-current (transverse) response formalism. We consider {\it hr}-{\it sB} phase in the pristine (undoped) form, however in the case of 8{\it Pmmn} the Fermi level is shifted 250 meV below the Dirac point to increase the carrier concentration. Figures \ref{fig5}(a) and \ref{fig5}(b) illustrate the plasmon dispersion of two phases of borophene along the two in-plane principal axes. The characteristics of the plasmon dispersion are seemed very much like in the typical 2D Dirac semimetal in the long-wavelength limit, showing $\sim\sqrt{\bf q}$ dependence. However, there is a crucial difference here, namely the very pronounced anisotropy in plasmon dispersion along the $\Gamma$X and $\Gamma$Y directions. This can be followed back to the strong anisotropy in the electronic structure\,\cite{silkin21} as well as to the anisotropy in the allowed optical (both intraband and interband) transitions along the two in-plane directions [as seen in Figs.\,\ref{fig3}(c) and \ref{fig3}(d)]. The similar behavior of 2D anisotropic plasmon dispersions has been reported in WTe$_2$\,\cite{torbatian2020tunable} and black phosphorous\,\cite{hyperBP3,hyperBP1,torbatian_PRB, ghamsari2020plasmon}. It should be pointed out that such strong anisotropy could lead to hyperbolic plasmons in 2D materials\,\cite{torbatian2020tunable}. Therefore, the combination of a high anisotropy and strong metallic character could make both phases of borophene promising 2D hyperbolic materials.

\begin{figure}[!t]
\begin{center}
\includegraphics[width=8.8cm]{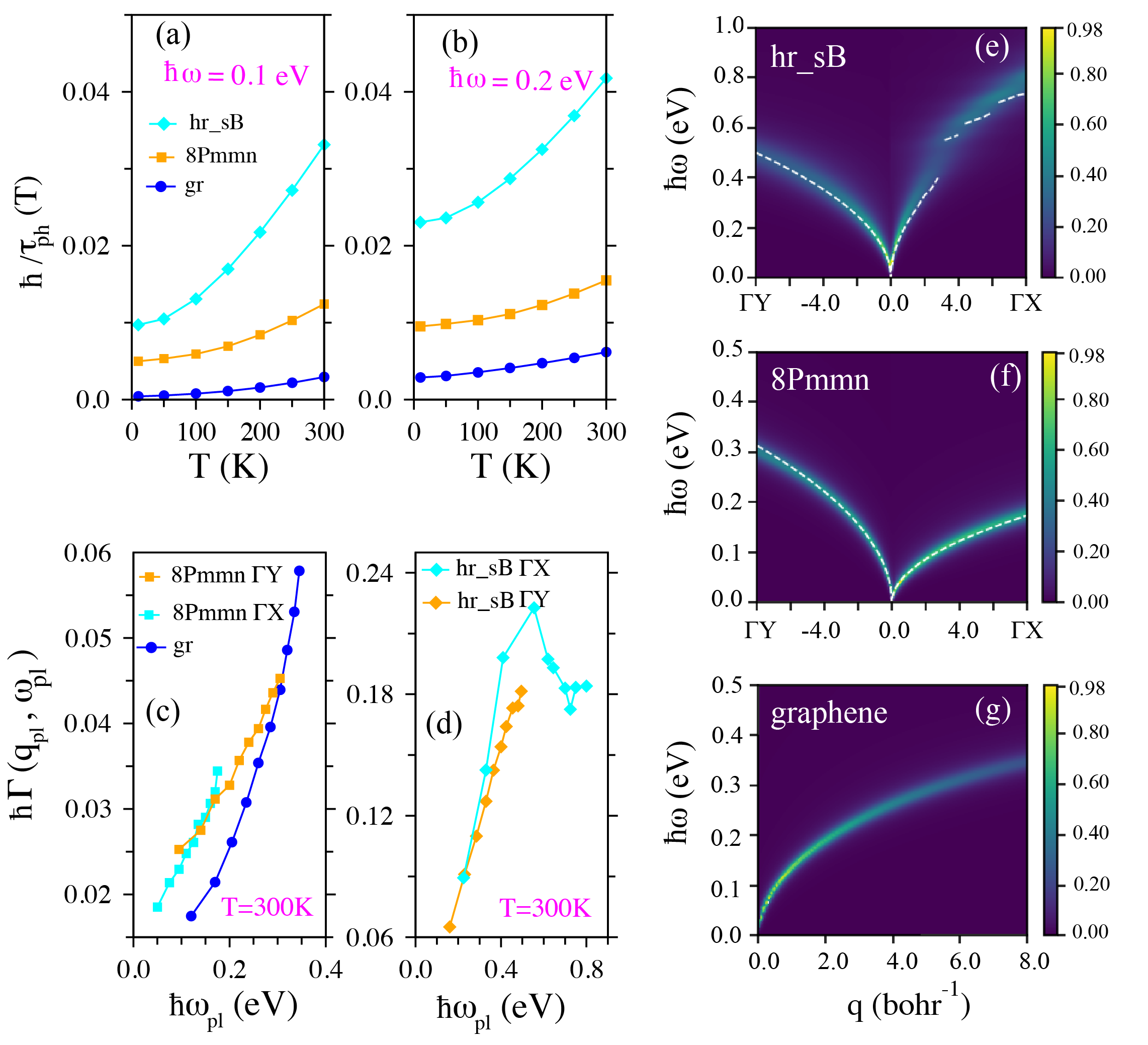}
\caption{{\bf Coupling of hyperbolic Dirac plasmons and phonons.} Plasmon damping rates owing to the electron-phonon coupling $\hbar/\tau_{\rm ph} (\omega)$ as a function of the temperature for {\it hr}-{\it sB}, 8{\it Pmmn} and graphene when energy is (a) $\hbar\omega=0.1$\,eV and (b) $\hbar\omega=0.2$\,eV. Total damping of plasmon which additionally includes Landau damping along the $x$ and $y$ directions and $T= 300$ for (c) graphene and 8{\it Pmmn} ($E_{\rm F}=-250$ meV) as well as for (d) {\it hr}-{\it sB}. Plasmon dispersions of (e) {\it hr}-{\it sB}, (f) 8{\it Pmmn}, and (g) graphene as a function of $q$ (in 10$^{-3}$ bohr$^{-1}$) when the electron-phonon coupling is included at $T= 300$\,K. The white dashed lines in (e) and (f) show the plasmon dispersions at $T= 0$\,K for {\it hr}-{\it sB} and 8{\it Pmmn}, respectively.
}
\label{fig6}
\end{center}
\end{figure}

In the following, we discuss the temperature dependence of the plasmon linewidth due to the EPC.
The case of hole-doped 8{\it Pmmn} borophene is additionally compared with the prototypical Dirac material graphene with $E_{\rm F}=-250$\,meV\,\cite{novko2020ultrafast}.
The damping rates $\hbar/\tau_{\rm ph}$ as a function of the temperature for two phases of borophene and graphene when $\hbar \omega =0.1$\,eV and $\hbar \omega =0.2$\,eV are plotted in Figs.\,\ref{fig6}(a) and \ref{fig6}(b). Compared to the 8{\it Pmmn} borophene and graphene, the {\it hr}-{\it sB} phase of borophene shows the largest damping rate due to the EPC and steepest temperature increase. For the presented plasmon energies, plasmon in graphene has the lowest damping rate. This is because under these conditions (i.e., $T\leq300$\,K and $\hbar\omega\leq 0.2$\,eV), graphene plasmon is only weakly coupled to acoustic phonons\,\cite{novko2020ultrafast}, while in both phases of borophene due to more complex structure there are multiple low-energy optical and acoustic phonons that can couple to Dirac plasmon [see Fig.\,\ref{fig2}].

To perform a more thorough survey of plasmon damping, we calculate the total damping of plasmon that additionally includes Landau damping (coupling to electron-hole pair transitions) by making use of  the following expression \cite{kupvcic2014damping}.
\begin{eqnarray}
\hbar\Gamma(\mathbf q,\omega)= 2\pi\hbar q {\rm Re}\{
\sigma_{\mu}(\mathbf q,\omega)\}
\label{dampeq}
\end{eqnarray}
where $\sigma_{\mu}(\mathbf q,\omega)$ contains both intraband $\sigma^{\rm intra}_{\mu}(\mathbf q,\omega)$ and interband $\sigma^{\rm inter}_{\mu}(\mathbf q,\omega)$ terms.
The total damping for all three systems is illustrated in Figs.\,\ref{fig6}(c) and \ref{fig6}(d). For two phases of borophene, $\hbar\Gamma(\mathbf{q}_{\rm pl},\omega_{\rm pl})$ is plotted along the $\Gamma$X and $\Gamma$Y directions. As expected, the total damping in the {\it hr}-{\it sB} phase of borophene is considerably larger than in the 8{\it Pmmn} phase and graphene, because both EPC and interband transitions are stronger in {\it hr}-{\it sB} [compare the intensities of ${\rm Re}\,\sigma_{\mu}$ in Figs.\,\ref{fig3}(c) and \ref{fig3}(d)].
Interestingly, the total plasmon damping of 8{\it Pmmn} and graphene are remarkably similar at $\hbar\omega_{\rm pl} \leq 0.3$\,eV. Furthermore, the total damping of Dirac plasmon in the {\it hr}-{\it sB} phase shows a specific sharp increase at $\sim 0.5$\,eV, followed by a decrease, but only in the $\Gamma$X direction. This interesting feature comes from the above-mentioed interband feature ($i$) that is only allowed for the $x$ polarization [compare ${\rm Re}\,\sigma_{x}(\omega)$ and ${\rm Re}\,\sigma_{y}(\omega)$ around 0.5\,eV in Figs.\,\ref{fig4}(b) and \ref{fig4}(e)]. Note that this coupling between Dirac plasmon and interband transitions ($i$) also results in avoided crossing in plasmon dispersion along the $\Gamma$X direction, as observed in Figs.\,\ref{fig5}(a) and \ref{fig6}(e).

\begin{figure*}[!t]
\begin{center}
\includegraphics[width=16.8cm]{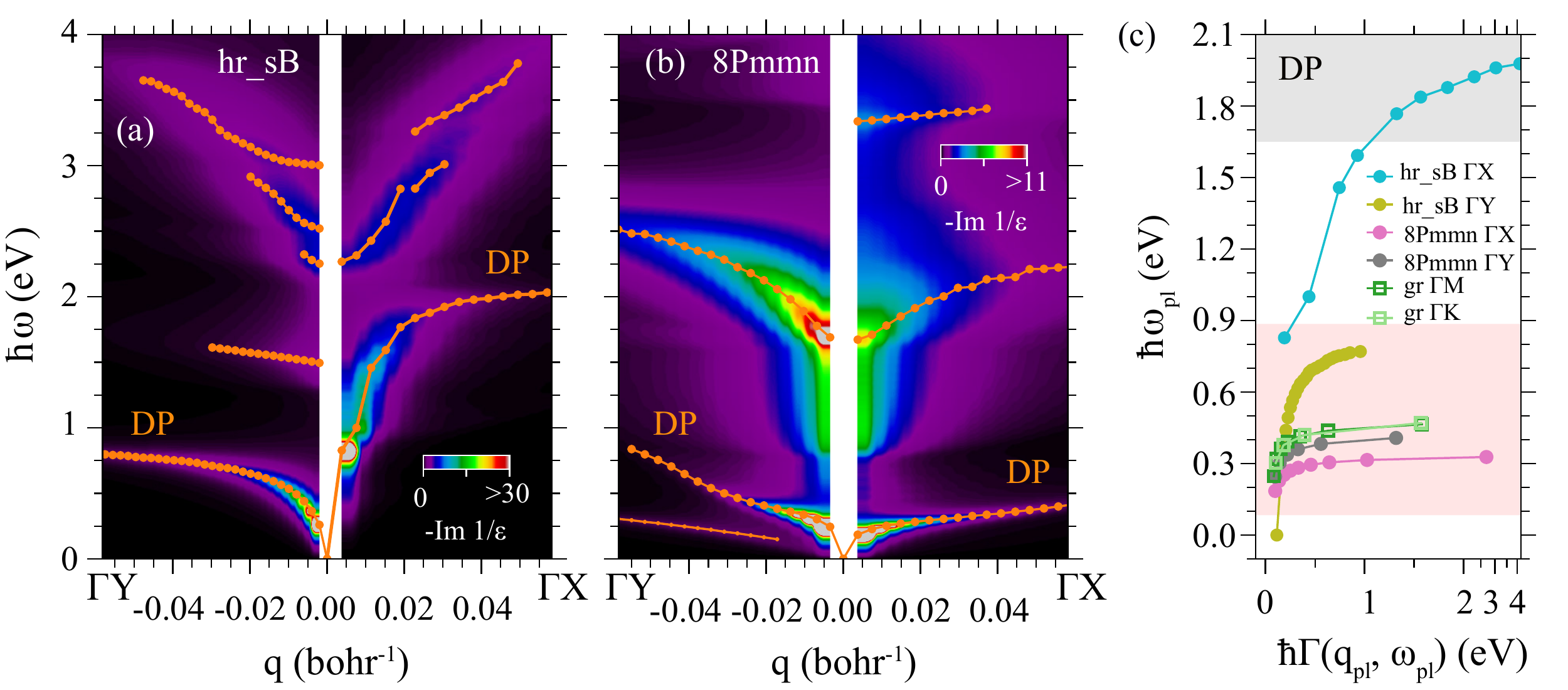}
\caption{{\bf Anisotropic electron excitations and damping in borophene.} The electron energy loss function ${\rm Im}[1/\varepsilon ({\bf q},\omega)]$ as a function of energy $\hbar\omega$ and momentum $q$ along the $\Gamma$X and $\Gamma$Y directions for (a) the pristine {\it hr}-{\it sB} and (b) the 8{\it Pmmn} ($E_{\rm F}=-250$\,meV) borophene. The orange dots follow the plasmon peaks in these two systems. Both phases show strongly anisotropic Dirac plasmon (DP) mode. (c) The total damping for two phases of borophene and graphene as a function of plasmon frequency $\hbar \omega_{\rm pl}$ and along the corresponding high symmetry directions. The mid-infrared and visible regions are highlighted with the red and grey shaded areas, respectively.
}
\label{fig7}
\end{center}
\end{figure*}

Figures \ref{fig6}(e)-(g) show the plasmon dispersions of pristine {\it hr}-{\it sB} as well as hole-doped 8{\it Pmmn} and graphene ($E_{\rm F}=-250$\,meV) along two principal axes at $T=300$\,K. The temperature dependence of the plasmon energy comes from the Fermi-Dirac distribution function and also from the EPC through $1/\tau_{\rm ph} (\omega)$ and $\lambda_{\rm ph} (\omega)$ functions. The temperature dependence of the plasmon dispersion due to the EPC is remarkable for the {\it hr}-{\it sB} borophene. On the other hand, the temperature dependence of the plasmon dispersion is weak in 8{\it Pmmn} and graphene owing to the weak coupling of plasmon and phonons in these two materials. To have an insight into the temperature dependence of the plasmon, we illustrate the plasmon dispersions at $T= 0$\,K for {\it hr}-{\it sB} and 8{\it Pmmn} with the white dashed line in Figs. \ref{fig6}(e) and (f) respectively. For 8{\it Pmmn}, it is obvious that the change in dispersion is negligible. On the other hand, it is considerable for {\it hr}-{\it sB}, especially along the $\Gamma$X where it changes from 0.73 eV to 0.80 eV at largest $q$ vector (8$\times$ 10$^{-3}$ bohr$^{-1}$) when temperature rises from $T= 0$\,K to $T=300$\,K.

Taking everything into account, these results show a impressive anisotropy in borophene of both plasmon dispersion and plasmon damping along the $\Gamma$X and $\Gamma$Y directions.

In addition, we examine the anisotropic electron excitons in the two phases of borophene in the short-wavelength regime (finite $\bf q$). To provide a dispersion relation of the plasmon modes in borophene when ${\bf q}>0$ we calculate the charge-charge response function and the corresponding dielectric function $\varepsilon({\bf q},\omega)$\,\cite{despoja2013two,torbatian2017plasmon,2018plasmonic}. The electron energy loss function ${\rm Im}[1/\varepsilon({\bf q},\omega)]$ for two phases of borophene is illustrated in Figs.\,\ref{fig7}(a) and \ref{fig7}(b) for $q\leq 0.06$\,bohr$^{-1}$ and $\omega\leq 4.0$\,eV. Note that for the smallest momentum the Dirac plasmon dispersion obtained by the charge-charge (longitudinal) response formalism is in good agreement with the results obtained by means of the current-current (transverse) response theory. For larger momenta, the remarkable anisotropy of Dirac plasmon in borophene is even more pronounced. Especially for the {\it hr}-{\it sB} phase, where plasmons energies, for instance, at $q=0.06$\,bohr$^{-1}$ are $\hbar\omega_{\rm pl}\approx 2$\,eV and $\hbar\omega_{\rm pl}\approx 0.8$\,eV along the $\Gamma$X and $\Gamma$Y directions, respectively. Also, the plasmon energy in the {\it hr}-{\it sB} phase reaches the visible range along the $\Gamma$X direction due to intrinsic high carrier density as well as the low intensity of interband transitions (and therefore minor interband screening) for the $x$ polarization [see Fig.\,\ref{fig3}(c)]. Relevant low intensity interband transitions are here interband continuum ($iii$) at around 1\,eV discussed above [see Figs.\,\ref{fig4}(c) and \ref{fig4}(d)]. On the other hand, strong interband feature ($ii$) at around 1.5\,eV screens and pushes the Dirac plasmon to lower energies for the $y$ polarization. 
For both phases of borophene, the interband screening makes the Dirac plasmon dispersionless and most importantly flat. This makes the group velocity of plasmon small, which might be important in the context of real-space localization of plasmon wave packets\,\cite{dajornada20}. Apart from the Dirac plasmons, here we can perceive the presence of the anisotropic screened interband excitations (i.e., interband plasmons). Particularly interesting are the interband excitations in the 8{\it Pmmn} phase that fall into the visible range, due to their high intensity and anisotropy. In addition, an acoustic plasmon is present in the 8{\it Pmmn} phase and it comes from the anisotropic Fermi velocity along two principal directions (i.e., anisotropic Dirac cone)\,\cite{pisarra14}.

Ultimately, we inspect the total plasmon damping for ${\bf q}>0$, which is shown in Fig.\,\ref{fig7}(c). It turns out that in the {\it hr}-{\it sB} phase the Dirac plasmon along the $\Gamma$X direction is strongly damped for energies in the visible range (i.e., $\hbar\Gamma>1$\,eV). The corresponding plasmon along the $\Gamma$Y direction, which falls into the mid-infrared region, is, on the other hand, moderately damped. In accordance with our conclusions for the plasmon damping in the long-wavelength limit, the plasmon decay rates of the hole-doped 8{\it Pmmn} borophene and graphene are similar. The unique considerable difference is the anisotropy of plasmon damping for the 8{\it Pmmn} borophene.

\section*{Conclusion}\label{sec:concl}
By using the density functional theory-based calculations we have examined optical properties and collective electron excitation modes (i.e., plasmons) in two phases of single-layer boron (borophene) that host Dirac fermions with tilted Dirac cone, namely the so-called {\it hr}-{\it sB} and 8{\it Pmmn} borophene. According to our numerical results, both of these phases are dynamically stable and are characterized with a strongly anisotropic electronic structure and optical absorption. In fact, hyperbolic plasmonic surfaces appear in the visible range. The {\it hr}-{\it sB} phase, with tilted Dirac cone and nodal line, supports strong hyperbolic Dirac plasmons already in the pristine form owing to the excessive concentration of carriers at the Fermi energy. The 8{\it Pmmn} borophene is a semimetal with Dirac point crossing the Fermi level, and therefore supports plasmon modes only upon carrier doping. For hole-doped 8{\it Pmmn} phase we have determined hyperbolic Dirac plasmons in the mid-infrared region. By analyzing the coupling of Dirac plasmon with phonons and electron-hole pair excitations, we show that the 8{\it Pmmn} phase has a very low plasmonic damping rate.

Both of these phases of borophene outperform typical two-dimensional Dirac material, i.e., graphene, as an optical and plasmonic device, owing to intrinsic metallic character, strong anisotropy, the appearance of hyperbolic plasmonic surfaces, low plasmon losses, and plasmons in the visible range.

\section*{METHODS}\label{sec:theo}
{\bf Computational details.--}
Our first-principles calculations are based on the density functional theory (DFT) within the QUANTUM ESPRESSO package 
\cite {0953-8984-21-39-395502} using Perdew-Zunger functional for exchange-correlation functional. The plane wave energy cutoff was fixed to $50$\,Ry. The 2D layers are modeled in periodic cells with $20$\,${\rm \AA}$ vacuum in the direction normal to the layer to reduce the interaction between periodic images. 
The convergence criterion for energy is set to 10$^{-8}$\,eV and the atomic positions are relaxed until the Hellmann-Feynman forces are less than 10$^{-4}$\,${\rm eV/\AA}$.
The 2D Brillouin zone (BZ) is sampled using $48 \times 24 \times 1$ and $30 \times 40 \times 1$ kpoint meshes for the {\it  hr}-{\it sB} and 8{\it Pmmn} phases of borophene, respectively. 
To accurately determine the electron-phonon coupling (EPC) constants, we employed recently developed Wannier interpolation technique\,\cite{wan90}, which has been implemented in the EPW code\,\cite{ponce2016epw}. To do so, the electronic structure, dynamical matrices, and electron-phonon matrix elements obtained from DFT and density functional perturbation theory (DFPT)\,\cite{baroni01} calculations and they were used as the initial data for Wannier interpolation within the maximally-localized-Wannier-functions formalism.

{\bf Optical absorption, phonon-induced damping, and electron excitation spectrum.--}
In this section, we present a short theoretical description of optical absorption quantities in the long-wavelength limit (${\mathbf q}\approx 0$) by making use of the current-current response tensor $\Pi_{\mu}(\mathbf q, \omega)$ calculated within DFT, where the electromagnetic interaction is mediated by the free-photon propagator. The thorough description of the method can be found in our previous works\,\cite{Novko2016,torbatian_PRB,torbatian2020low}.

The optical conductivity can be calculated as $\sigma_{\mu}(\omega)=-i\lim_{\mathbf{q}\rightarrow0}\Pi^0_{\mu}(\mathbf q, \omega)/\omega$, while the optical absorption is given by $A(\mathbf q,\omega)=-4\hbar {\rm Im}\,\Pi_{\mu}(\mathbf{q},\omega)/\omega$\,\cite{Novko2016,novko17,torbatian_PRB}, where $\Pi^0_{\mu}(\mathbf q, \omega)$ is a bare current-current tensor, while $\Pi_{\mu}(\mathbf q, \omega)$ is a fully screened tensor.

To investigate the effects of phonons on the plasmon dispersion, we impliment the formalism presented in Refs.\,\cite{novko17,Caruso2018}. Optical excitations are first convenient to decompose into the intraband ($n=m$) and interband ($n\ne m$) contributions, i.e., $\Pi^{\rm 0}_{\mu}=\Pi^{\rm 0,intra}_{\mu}+\Pi^{\rm 0,inter}_{\mu}$. The electron-phonon scattering mechanism is then considered in the intraband channel.

In the optical limit when ${\mathbf q}\approx 0$, the non-interacting interband current-current response tensor is
\begin{align}
\Pi^{\rm 0,inter}_{\mu}(\omega)= \frac{2}{V} \sum_{\mathbf k, n\neq m} \frac{\hbar \omega}{E_n(\mathbf k)-E_m(\mathbf k)}\nonumber\\
\times \left|J^\mu_{nm\mathbf k}\right|^2\frac{f_n(\mathbf k)-f_m(\mathbf k)}{\hbar\omega + i\eta+ E_n(\mathbf k) - E_m(\mathbf k)},
\label{current}
\end{align}
where $J^\mu_{nm\mathbf k}$ are the current vertices, $E_n(\mathbf k)$ are the Kohn-Sham energies, $f_n(\mathbf k) $ is the Fermi-Dirac distribution function at temperature $T$, $\mu$ are the polarization directions, and $V$ is the normalized volume. Further, the summation over $\mathbf k$ wavevectors is carried on a 200$\times$70$\times$1 (180$\times$240$\times$1) grid for {\it hr}-{\it sB} (8{\it Pmmn}) phase of borophene, $n$ band index sums over $30$ electronic bands.

The corresponding intraband contribution of current-current response tensor can be written as the following~\cite{novko17,PhysRevB.3.305}:
\begin{align}
\Pi^{\rm 0,intra}_{\mu}(\omega)= \frac{2}{V} \frac{\omega}{\omega[1+\lambda_{\rm ph}(\omega)]+i/\tau_{\rm ph}(\omega)}\sum_{\mathbf k, n} \frac{\partial f_{nk}}{\partial E_{n\mathbf k}}  |J^\mu_{nn\mathbf k}|^2.
\label{intra_cu}
\end{align}
Here the effects of the EPC are contained in the temperature-dependent dynamical scattering time and energy renormalization parameters, i.e., $\tau_{\rm ph}(\omega)$ and $\lambda_{\rm ph}(\omega)$, respectively. The dynamical scattering time is given by  \cite{novko2020broken,novko2018nonadiabatic}
\begin{align}
\hbar/\tau_{\rm ph}(\omega)=\frac{\pi\hbar}{\omega} \int d\Omega \alpha^2F(\Omega)\Big[2\omega \coth\frac{\Omega}{2k_BT}\nonumber\\-(\omega+\Omega)\coth\frac{\omega+\Omega}{2k_BT}+(\omega-\Omega)\coth\frac{\omega-\Omega}{2k_BT}\Big]
\label{tau}
\end{align}
where $k_B$ is the Boltzmann constant and
$\alpha^2F(\Omega)$ is the Eliashberg spectral function\,\cite{novko17,Caruso2018} 
\begin{equation}
\label{eq:spectral}
\alpha^2F(\omega)=\frac{1}{2\pi N(e_F)}\sum_{{\bf q}\nu}\delta(\omega-\omega_{{\bf q}\nu})\frac{\gamma_{{\bf q}\nu}}{\hbar\omega_{{\bf q}\nu}},
\end{equation}
in which $\omega_{{\bf q}\nu}$ and $\gamma_{{\bf q}\nu}$ are the frequency and linewidth for phonon mode $\nu$ at wavevector ${\bf q}$,
$N(e_F)$ is the density of states at the Fermi level.
Finally, the dynamical energy renormalization parameter $\lambda_{\rm ph}(\omega)$ is obtained by performing the Kramers-Kronig transformation of $1/\tau_{\rm ph}(\omega)$.

The total EPC constant can be determined by either Brillouin-zone summation or the frequency-space integration.
\begin{equation}
\label{eq:lambda}
\lambda=\sum_{{\bf q}\nu}\lambda_{{\bf q}\nu}=2\int\frac{\alpha^2F(\omega)}{\omega}d\omega,
\end{equation}
where $\lambda_{{\bf q}\nu}$ is the mode- and momentum-resolved EPC constant. Notice that the phonon properties (i.e., phonon energies and electron-phonon matrix elements), needed for calculating $\alpha^2F(\omega)$ are obtained within the EPW code\,\cite{ponce2016epw}. The calculations of the EPC constants and Eliashberg function are performed on dense grids of 400$\times$200 (120$\times$160) k- and 80$\times$40 (30$\times$40) q-points for {\it hr}-{\it sB} (8{\it Pmmn})  phase of borophene, which ensures the numerical convergence of the results presented in this work. Having considered these results, we can calculate scattering rate by using Eq.\,\eqref{tau} and finally the phonon-induced plasmon decay for both phases of borophene.

The electronic excitation spectra within the random-phase approximation (RPA) for finite momentum (${\bf q}>0$) are calculated by means of charge-charge correlation function formalism. The dielectric function in the RPA is given by
\begin{equation}
\varepsilon_{\textbf{G}\textbf{G}'}(\textbf{q},\omega)=
\delta_{\textbf{G}\textbf{G}'}-
\sum_{\textbf{G}_{1}}v_{\textbf{G}\textbf{G}_{1}}(\textbf{q})\chi^{0}_{\textbf{G}_{1}\textbf{G}'}(\textbf{q},\omega),
\label{eq:rpa1}
\end{equation}
where the charge-charge correlation function is given by
\begin{eqnarray}
&&\chi^{0}_{{\bf G}{\bf G}'}({\bf q},\omega)=\frac{2}{V}\sum\limits_{{\bf k},nm}\ \frac{f_{n}({\bf k})-f_{m}({\bf k}+{\bf q})}{\hbar\omega+i\eta+E_{n}({\bf k})-E_{m}({\bf k}+{\bf q})} \nonumber \\
&&\times M_{n{\bf k},m{\bf k}+{\bf q}}({\bf G})\ M^*_{n{\bf k},m{\bf k}+{\bf q}}({\bf G'}).
\label{eq:rpa2}
\end{eqnarray}
Here ${\bf G}$ is the reciprocal lattice wave vector and $v_{\textbf{G}\textbf{G}'}$ is the bare Coulomb interaction. The charge matrix elements are defined as  
\begin{equation}
M_{n{\bf k},m{\bf k}+{\bf q}}({\bf G})=\left\langle \psi_{n{\bf k}}\left|e^{-i({\bf q}+{\bf G}){\bf r}}\right|\psi_{m{\bf k}+{\bf q}}\right\rangle_V,
\label{eq:rpa3}
\end{equation}
where $\psi_{n{\bf k}}(\mathbf{r})$ are Kohn-Sham wave functions. The electron energy loss spectrum is obtained by calculating
\begin{equation}
\mathrm{EELS}=-\mathrm{Im}\left[1/\varepsilon_{\mathbf{GG'}}(\mathbf{q},\omega)\right]_{\mathbf{G=G'=0}},
\label{eq:rpa4}
\end{equation}
which, in addition, contains plasmon poles at $\hbar\omega_{\rm pl}$ as well as the screened interband and intraband excitations. The corresponding calculations are performed within the GPAW\,\cite{gpaw} code using the PAW pseudopotentials with energy cutoff of 600 eV, LDA functional, and a $30\times20\times1$ ($20\times30\times1$) electron-momentum grid for {\it hr}-{\it sB} (8{\it Pmmn}). The denser k grids are used for the summations performed in Eq.\,\eqref{eq:rpa2}, namely $300\times200\times1$ and $200\times300\times1$ for {\it hr}-{\it sB} and 8{\it Pmmn}, respectively. We include up to 30 electronic bands, $\eta=30$ meV for the broadening parameter, and the cutoff is 10\,eV for the ${\bf G}$ vectors. Notice we are convinced that ground state properties (i.e., electronic band structure) obtained from the two separate DFT codes are practically identical.

\begin{acknowledgments}
D.N. acknowledges financial support from the Croatian Science Foundation (Grant no. UIP-2019-04-6869) and from the European Regional Development Fund for the ``Center of Excellence for Advanced Materials and Sensing Devices'' (Grant No. KK.01.1.1.01.0001). R. A. was supported by the Australian Research Council Centre of Excellence in Future Low-Energy Electronics Technologies (project number CE170100039).
\end{acknowledgments}

\bibliography{B}

\end{document}